# A Novel Adaptive Hybrid Focal-Entropy Loss for Enhancing Diabetic Retinopathy Detection Using Convolutional Neural Networks

Santhosh Malarvannan
*School of Computer Science and Engineering*
*Vellore Institute of Technology, Chennai*
Tamil Nadu, India
sandy5501m@gmail.com

Pandiyaraju V
*School of Computer Science and Engineering*
*Vellore Institute of Technology, Chennai*
Tamil Nadu, India
pandiyaraju.v@vit.ac.in

Shravan Venkatraman
*School of Computer Science and Engineering*
*Vellore Institute of Technology, Chennai*
Tamil Nadu, India
shravan.venkatraman18@gmail.com

Abeshek A
*School of Computer Science and Engineering*
*Vellore Institute of Technology, Chennai*
Tamil Nadu, India
abeshek.a2021@vitstudent.ac.in

Priyadarshini B
*School of Computer Science and Engineering*
*Vellore Institute of Technology, Chennai*
Tamil Nadu, India
priyadarshinibala01@gmail.com

Kannan A
*Department of Information Science and Technology*
*College of Engineering, Guindy, Anna University, Chennai*
Tamil Nadu, India
akannan123@gmail.com

*Abstract*— Diabetic retinopathy is a leading cause of blindness around the world and demands precise AI-based diagnostic tools. Traditional loss functions in multi-class classification, such as Categorical Cross-Entropy (CCE), are very common but break down with class imbalance, especially in cases with inherently challenging or overlapping classes, which leads to biased and less sensitive models. Since a heavy imbalance exists in the number of examples for higher severity stage 4 diabetic retinopathy, etc., classes compared to those very early stages like class 0, achieving class balance is key. For this purpose, we propose the Adaptive Hybrid Focal-Entropy Loss which combines the ideas of focal loss and entropy loss with adaptive weighting in order to focus on minority classes and highlight the challenging samples. The state-of-the art models applied for diabetic retinopathy detection with AHFE revealed good performance improvements, indicating the top performances of ResNet50 at 99.79%, DenseNet121 at 98.86%, Xception at 98.92%, MobileNetV2 at 97.84%, and InceptionV3 at 93.62% accuracy. This sheds light into how AHFE promotes enhancement in AI-driven diagnostics for complex and imbalanced medical datasets.

## I. INTRODUCTION

The application of Artificial Intelligence (AI) technology in the field of medical imaging has witnessed tremendous progress over the past few years, changing the way medical practitioners detect and manage diseases in different parts of the world. Since the rise of Deep learning networks and in particular Convolutional neural networks (CNNs), the medical image interpretation process has been automated and it has become more accurate and faster than the conventional methods. AI models can detect patterns in images that, if looked at by the human eye, might seem trivial and make it more accurate when diagnosing a patient. For example, a recent study by Sharmiladevi et al. [1] claims that patients can avoid manual inspection of a patient for skin cancer using AI tools which helps both in the accuracy of diagnosis and time/cost for inspections. With these advancements, the diagnosis of diabetic retinopathy (DR) is also becoming possible, where the AI can classify the stages of the disease from photographs of the retina.

Diabetic retinopathy (DR) occurs in approximately 30 percent of people with diabetes and remains one of the most significant complications associated with the condition. Diabetes related complications global in scale and lead to substantial vision loss and in many cases blindness [2]. Millions of people across the globe are affected by DR and this adds sweeping challenges to healthcare systems especially in areas with a rising diabetic population base. Gary et al. (2017) further pointed out the flaws in traditional manual detection and treatment methods for DR, arguing that such approaches are not only costly but time consuming and usually involve an element of human mistake. For this reason, Convolutional Neural Networks (CNNs) have become very helpful in automating DR diagnosis because they drastically enhance the diagnostic accuracy [3]. The network scans the retina for the presence of microaneurysms, hemorrhages, exudates, and so forth. Despite their success, there is one other proximate problem with using CNN models which still persists; these models exhibit a consistent weakness to imbalanced data of which is a standard feature in medical datasets. This even distribution of cases where normal cases far outnumber the abnormal cases tends to distort the model's performance and as such, detection rates for classes that are underrepresented are suboptimal. Therefore, while CNNs offer a promising solution for effective DR detection, more work has to be done to address the issues of imbalanced data for these approaches to be effective in practice.

In the last few decades, diabetes has emerged as a pandemic with unprecedented speed, as it was estimated that there are 382 million diabetes patients worldwide, a number which is expected to rise to 592 million patients by 2025. Diabetic retinopathy, which is the most common complication of diabetes, affects 34.6% of diabetic patients, and this condition involves the mutilation of retinal blood vessels. This commonly unnoticed loss of vision which occurs insidiously in the initial phases has made the need for accurate detection much earlier than the later stages of disease management when prognosis of vision is nearly impossible [4]. There is also a significant proportion of patients at risk of vision threatening complications, as 7% of patients go on to develop proliferative diabetic retinopathy and 6.8% develop diabetic macular edema(DME). The above statistics demonstrate the inadequacy of present diagnostic techniques to detect and categorize diabetic retinopathy at earlier stages before they advance to the proliferative or moderate stages [5].

Despite the advancement in diagnostic methods, most of these still necessitate manual assessment, which is inefficient, subjective and inconsistent in quality. This is quite worrying in countries where the incidence of diabetes is high like in the Khyber Pakhtunkhwa province of Pakistan, where around 30% of the population is diabetic and 4% of blindness cases are attributed to DR. As manual approaches fail to meet the growing demand, automated strategies using deep learning techniques, such as Convolutional Neural Networks (CNNs) seem to be an efficient and more scalable solution [6]. However, issues such as the examinations being performed on unbalanced medical images, where the abnormal cases are few, still prevent the best performance, and thus, more effective strategies need to be developed.

In this study, a new technique is proposed that combine the present trend of training deep learning models with the traditional methods of handling data imbalance in the task of diabetic retinopathy detection. The AHFE formulation is instituted to improve the model orientation towards learning invariant features while simultaneously addressing class imbalance. Even though only mild and moderate lesions are tolerable in the image for early diabetic retinopathy diagnosis, the loss function can also be trained on datasets with greater severity. The scope of AHFE allows the model to learn with reduced overfitting on the majority class while learning about the hard examples of the majority class through the focal loss mechanism. The fusion of these forces creates an ideal situation in which stochastic gradient descent-related phenomena can increase the sensitivity of the model and improve overall performance. The collected evidence confirms our hypothesis that links the incorporation of the AHFE loss function with an increase in the accuracy of CNN models. This increases the reliability of the model in diabetic neuropathy detection and demonstrates the ease of overcoming data imbalance and providing practical techniques for diagnosis that can be implemented efficiently.

## II. RELATED WORKS

Shamrat et al. [7] introduced DRNet13, a novel convolutional neural network designed to automate diabetic retinopathy staging. Trained on an augmented dataset of 7,500 fundus images, the model achieved a 97% accuracy rate, surpassing fifteen pre-trained models in both speed and efficiency. However, some misclassifications were noted, and reliance on a single dataset may limit the generalizability of the findings. Thanikachalam et al. [8] developed a Deep Convolutional Neural Network (CNN) model for the automatic detection and classification of Diabetic Retinopathy (DR) and Diabetic Macular Edema (DME). Their approach incorporated Discrete Wavelet Transform for image preprocessing and Adaptive Gabor Filtering for feature extraction. The CNN classifier's performance was enhanced using the Chicken Swarm Algorithm, resulting in an accuracy of 97.91%. Despite this achievement of high accuracy, the study's reliance on a single dataset may limit the generalizability of the findings. Qummar et al. [9] developed an ensemble model combining five CNN architectures—ResNet50, InceptionV3, Xception, DenseNet121, and DenseNet169—to detect all stages of diabetic retinopathy, emphasizing early detection. Trained on the Kaggle fundus image database, the ensemble achieved a validation accuracy of 80.8%, outperforming individual models. However, the study's reliance on a single dataset may limit the generalizability of its findings. Dutta et al. [10] proposed an automated model for diabetic retinopathy detection using Backpropagation Neural Networks (NN), Deep Neural Networks (DNN), and Convolutional Neural Networks (CNN). The model employed weighted Fuzzy C-means for class severity classification based on fundus image features. Results showed that DNN outperformed CNN in training and validation accuracy due to CPU limitations affecting CNN performance. However, noisy images reduced prediction accuracy, and future improvements include GPU-based training and integration with existing screening algorithms for enhanced efficiency. Lam et al. [11] utilized Convolutional Neural Networks (CNNs) on color fundus images to stage diabetic retinopathy, achieving high sensitivity in binary classification. They noted challenges in detecting subtle features in early stages and suggested preprocessing and expert verification to enhance accuracy. Additionally, they proposed a two-step lesion detection approach using shallow CNNs to improve detection of less prominent disease features. However, specific performance metrics were not reported, and reliance on a single dataset may limit the generalizability of their findings.

Tymchenko et al. [12] introduced a single-image fully automated deep learning method for diabetic retinopathy stage detection. They utilized an ensemble of three different CNN architectures with transfer learning and attained 0.99 sensitivity and specificity on the APTOS 2019 dataset. Their approach successfully minimized variance and improved generalization. They also proposed that adding SHAP analyses and meta-learning methods would further enhance model performance. Khan et al. [13] introduced the VGG-NIN model, integrating VGG16 with spatial pyramid pooling and a network-in-network architecture to efficiently classify diabetic retinopathy stages. This design reduces computational complexity by minimizing learnable parameters while maintaining high accuracy. The model achieved an accuracy of 85% in four-class DR grading using the Kaggle dataset. Nonetheless, its performance may vary with different datasets, indicating a need for further validation to assess its generalizability. Fayyaz et al. [14] developed a deep learning framework utilizing AlexNet and ResNet101 for feature extraction, combined with Ant Colony Optimization for feature selection and a multi-kernel SVM for classifying diabetic retinopathy severity. This approach achieved a 93% accuracy in classifying fundus images. The study, however, did not extensively discuss potential limitations or challenges, such as computational complexity or generalizability across diverse datasets. Chakrabarty [15] proposed a deep learning approach for diabetic retinopathy diagnosis using Convolutional Neural Networks (CNNs). The methodology involved preprocessing fundus images by converting them to greyscale and resizing, followed by classification using a CNN. This approach achieved high accuracy, effectively reducing doctors' workload through early and accurate disease detection.

Özbay [16] proposed a new Active Deep Learning (ADL) model for diagnosing diabetic retinopathy through segmentation with multi-layer CNN and artificial bee colony (ABC) algorithm placement. The model sorts the clinical DR lesions into five severity levels with an accuracy rate of 99.66% on the EyePacs benchmark dataset. This method exemplifies the efficacy of deep learning techniques applied after images have undergone pre-processing. Nahiduzzaman et al. [17] introduced a hybrid model combining Convolutional Neural Networks (CNN) and Singular Value Decomposition (SVD) for feature extraction, alongside an Extreme Learning Machine (ELM) for classifying diabetic retinopathy (DR) stages. Utilizing Ben Graham's preprocessing and Contrast Limited Adaptive Histogram Equalization (CLAHE), the model achieved 99.73% precision and 100% sensitivity in binary classification, and 98.09% accuracy for five-stage DR classification on the APTOS-2019 dataset. However, the model's performance on the Messidor-2 dataset was slightly lower, indicating potential variability across different datasets. Daanouni et al. [18] introduced the NSL-MHA-CNN model, enhancing MobileNet's robustness against adversarial attacks by integrating Neural Structure Learning (NSL) and Multi-Head Attention (MHA). This approach maintained 98% accuracy under perturbations, ensuring reliable and cost-effective diabetic retinopathy diagnosis. However, the model's performance in real-world clinical settings with diverse data sources remains to be validated.

## III. METHODOLOGY

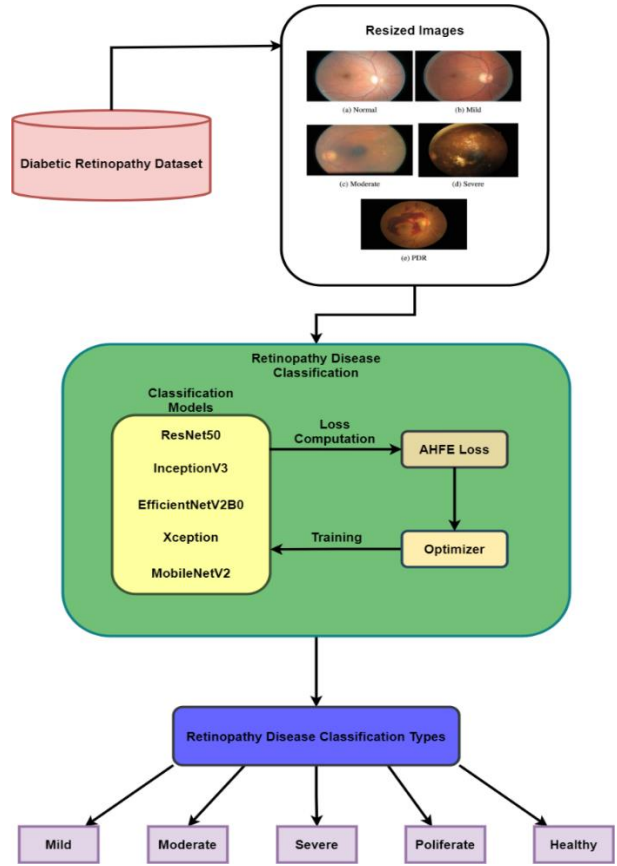

Figure 1. Overall Proposed Architecture for Diabetic Retinopathy Detection

### A. Dataset Information

The APTOS 2019 dataset refers to the baseline data for the aspirants of the APTOS 2019 Blindness Detection competition which was conducted by Asia Pacific Tele-Ophthalmology Society (APTOS). The dataset of over 3,662 retinal fundus images with accompanying DR grade has been provided and categorized as 0 (No DR), 1 (Mild DR), 2 (Moderate DR), 3 (Severe DR) and 4 (Proliferative DR). Considering the images include real-life environments making this dataset beneficial for ML as the models need these images for the development of DR detection and grading systems. The dataset also shows class improvement and is more appropriate for analyzing the various DR stages and enhancing the diagnosis of vision loss related disorders.

### B. Dataset exploration

This data exploration includes images from APTOS 2019. The same images are always used for the diagnosis of Diabetic Retinopathy. Corresponding to the two retinal images, the five classes of Diabetic Retinopathy are listed below, from Class 0 to Class 4:

**Class 0 (No DR):** No visible signs of diabetic retinopathy; clear retina with no lesions.

**Class 1 (Mild DR):** Few scattered microaneurysms indicating early-stage DR.

**Class 2 (Moderate DR):** Increased microaneurysms and small hemorrhages, suggesting progression.

**Class 3 (Severe DR):** Numerous hemorrhages and neovascularization, indicating severe retinal damage.

**Class 4 (Proliferative DR):** Extensive neovascularization and retinal scarring, with a high risk of vision loss.

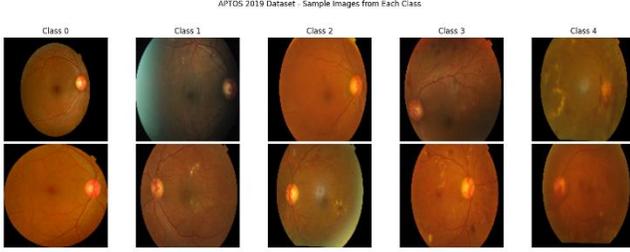

Figure 2. Sample images from the APTOS 2018 dataset

Figure 2 provides a quick overview of the progression of retinal changes across different severity levels, helping to illustrate the features that differentiate each class.

## IV. PROPOSED WORK

In this research, we propose a novel loss function, named as the Adaptive Hybrid Focal-Entropy (AHFE) Loss, designed to overcome certain challenges associated with the Categorical Cross-Entropy (CCE) Loss. CCE is one of the standard losses in multi-class classification which works on the principle of the probability for each predicted class and its true class by matching them and penalizing wrong predictions more. It calculates the loss function of negative log likelihood for the predicted probability of the true class, which is higher when the prediction for his class differs more. It is clear that CCE increases with the number of classes and reduces diversity in predictions since these values are summed up for every class and normalised over the batch during training.

For a batch of N samples, the total CCE Loss is averaged as:

$$CCE_{total} = -\frac{1}{N}\sum_{i=1}^{N}\sum_{k=1}^{N} y_{ik} \log(p_{ik}) \quad (1)$$

However, its performance gets poor where there is class imbalance, or difficult or overlapping classes. AHFE utilizes both entropy loss and focal loss to improve the robustness and adaptability of the proposed approach, especially for its application in diabetic retinopathy detection. Here, we bring out some of the main shortcomings associated with CCE and the solution provided by AHFE.

**i. Simultaneous Hard-Sample Emphasis and Confidence Control**
CCE is quite passive when it comes to identifying those hard-to-classify samples and tends to be lagging when it comes to rectifying such instances. AHFE employs focal loss principles which de-emphasize easy samples while assigning greater weight to more challenging ones. Additionally, it incorporates entropy-based regularization to mitigate overconfidence, preventing the model from overfitting to dominant patterns. This dual mechanism ensures optimal learning on complex patterns thus resulting in higher accuracy and better generalization on difficult datasets.

**ii. Adaptive Learning Adjustment**
CCE applies a fixed loss formulation, ignoring the need for adaptation during training. However, AHFE dynamically adjusts its entropy and focal terms, enabling it to shift focus from simpler to more complex samples as training progresses. This adaptive mechanism promotes balanced generalization in all classes, making AHFE more effective than traditional loss functions that rely on static weighting strategies.

**iv. Reduction of Overfitting on the Majority Class**
CCE often leads to overfitting, favouring majority classes and reducing sensitivity to minority classes. AHFE addresses this by combining adaptive weighting, focal loss, and an entropy-driven term to prevent the model from excessively focusing on dominant classes. This approach enhances recall for underrepresented classes without compromising generalization, which is essential in medical settings where missing rare cases can be costly.

**vi. Mitigation of Noisy Labels and Class Overlap with a Hybrid Loss Approach**
While Categorical Cross-Entropy (CCE) is generally effective for generalization, it struggles with noisy labels and overlapping classes. In contrast, AHFE combines a focal component to reduce the impact of noisy samples with entropy adjustments that enhance discrimination among overlapping classes. This dynamic adaptation to class overlap and uncertainty makes AHFE particularly advantageous for complex datasets.

The main designed loss function, Adaptive Hybrid Focal-Entropy (AHFE) was proposed with the intention of addressing umbilical challenges Categorical Cross-Entropy (CCE) possesses with importance in scenarios involving complexities of data structures like unbalanced class structure, certain difficulty of samples and even noise in labels. In direct contrast to CCE, which treats all the classes and samples equally regardless of their functions or purposes; AHFE employs both focal loss and adaptive entropy as its prime purposes and offers a more balanced approach. This dual-focused design enhances

AHFE's capacity to handle intricate class structures, focusing on the minority classes and difficult instances while maintaining generalization.

1. For the Focal Loss component which focuses on samples which are difficult to classify, reducing the impact of the easily classified ones,

$$FL(i) = -\frac{1}{N}\sum_{i=1}^{N}(1-p_{ik})^{\gamma} y_{ik}\log(p_{ik}) \quad (2)$$

Where:
- $\gamma$ denotes the tunable focusing parameter
- $p_{ik}$ denotes the probability of the correct class for each sample
- $y_{ik}$ denotes the ground truth indicator

2. The Entropy Loss component helps penalize uncertain predictions:

$$EL(i) = -\sum_{i=1}^{N} p_{ik}\log(p_{ik}) \quad (3)$$

3. Adaptive weights $\alpha_k$ are calculated depending on class frequency to prioritize minority classes:

$$\alpha_k = \frac{1}{\sqrt{N_k + \epsilon}} \quad (4)$$

Where:
- $N_k$ denotes the number of samples in class k
- $\epsilon$ denotes a small constant

4. Full AHFE Loss Function
Combining Focal Loss and Entropy Loss along with adaptive weights and a balancing parameter $\lambda$, for a batch of N samples:

$$AHFE_{total} = -\frac{1}{N}\sum_{i=1}^{N}\sum_{k=1}^{N}\alpha_k[(1-p_{ik})^{\gamma} y_{ik}\log(p_{ik}) + \lambda p_{ik}\log(p_{ik})] \quad (5)$$

AHFE, on the other hand, addresses this problem by modifying the weighting mechanisms relative to the sample and class distribution on the fly, which results in paying more attention to cases that are frequently ignored by classical loss functions. As it is capable of further adaption, it also corrects the biases, decreases the overfitting on the main classes, and remains tolerant to noisy labels, which is the scope of the problems faced by CCE to a great extent. Therefore, AHFE is effective in complex and unequally represented datasets to enhance classification performance, which is a remarkable development in the area of healthcare and other domains involving rare-event detection requiring specific and accurate classification.

## V. RESULTS AND DISCUSSION

This paper proposes a novel loss function for deep-learning frameworks. For our experiment, we have opted to implement our loss function for the classification diabetic retinopathy diseases. We have selected five existing deep learning models to evaluate the performance of the proposed loss function. These deep learning models were trained for 128 epochs, and the performance metrics of these models were recorded and plotted for each epoch.

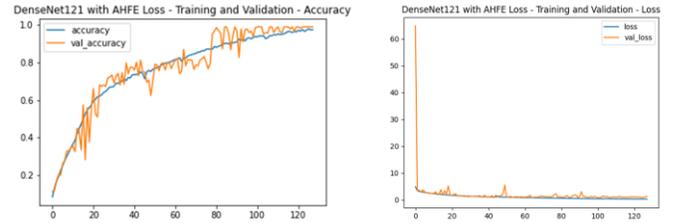

Figure 3. Training and validation plot for DenseNet121

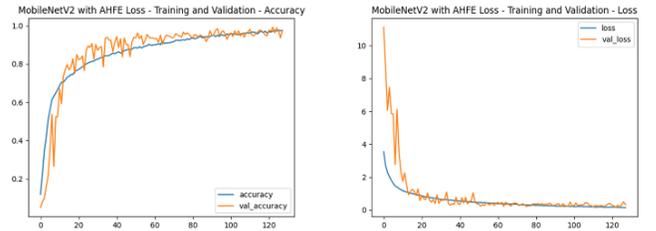

Figure 4. Training and validation plot for MobileNetV2

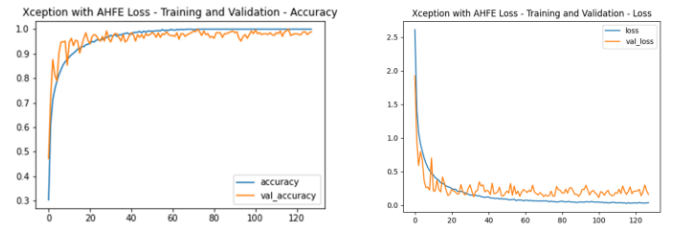

Figure 5. Training and validation plot for Xception

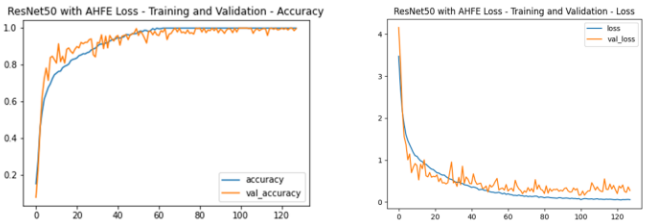

Figure 6. Training and validation plot for ResNet50

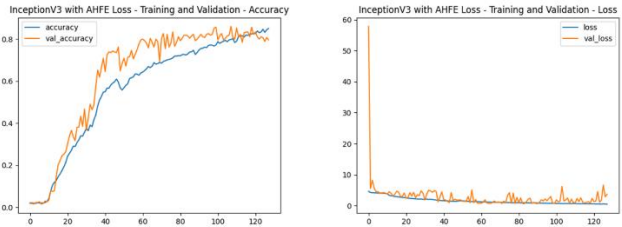

Figure 7. Training and validation plot for InceptionV3

To evaluate the effectiveness of the Adaptive Hybrid Focal-Entropy (AHFE) Loss, we trained five state-of-the-art (SOTA) deep learning models—ResNet50, DenseNet121, Xception, MobileNetV2, and InceptionV3—on the APTOS dataset and compared their performance using Categorical Cross-Entropy (CCE) Loss and AHFE Loss (Tables 1 and 2).

**Categorical Crossentropy Loss Function**

| Model | Accuracy (%) | Precision (%) | Recall (%) | F1 - Score (%) |
|---|---|---|---|---|
| DenseNet121 | 90.57 | 90.26 | 90.17 | 90.21 |
| ResNet50 | 91.50 | 90.56 | 90.72 | 90.63 |
| Xception | 92.54 | 92.15 | 92.76 | 92.45 |
| MobileNetV2 | 90.15 | 90.57 | 90.24 | 90.40 |
| InceptionV3 | 82.64 | 82.06 | 81.65 | 81.85 |

Table 1. Performance evaluation on the APTOS Dataset using Categorical Crossentropy Loss Function

**AHFE Loss Function**

| Model | Accuracy (%) | Precision (%) | Recall (%) | F1 - Score (%) |
|---|---|---|---|---|
| DenseNet121 | 98.86 | 97.53 | 98.33 | 97.93 |
| **ResNet50** | **99.76** | **98.57** | **99.29** | **98.93** |
| Xception | 98.92 | 98.26 | 97.71 | 97.98 |
| MobileNetV2 | 97.84 | 96.96 | 97.37 | 97.16 |
| InceptionV3 | 93.62 | 92.94 | 93.17 | 93.05 |

Table 2. Performance evaluation on the APTOS Dataset using AHFE Loss Function

From the above tables 1 and 2, we can conclude that a consistent improvement in accuracy, precision, recall, and F1-score can be observed across all models with AHFE Loss. ResNet50 achieved the highest accuracy at 99.76%, significantly surpassing its CCE-based counterpart. Similarly, Xception and DenseNet121 improved to 98.92% and 98.86%, respectively, highlighting AHFE's role in enhancing model sensitivity and class balance. Compared to CCE Loss, which struggles with imbalanced data and overfitting to dominant classes, AHFE integrates focal loss principles to emphasize hard-to-classify samples and entropy regularization to prevent overconfidence, leading to more stable learning. Notably, InceptionV3, which underperformed with CCE (82.64%), improved to 93.62% with AHFE, reinforcing its effectiveness in handling challenging datasets like diabetic retinopathy. Beyond outperforming CCE, AHFE surpasses conventional loss functions like Focal Loss and Class-Balanced Loss by dynamically adjusting weights based on sample difficulty and class distribution, ensuring better generalization. These findings confirm AHFE as a superior alternative for diabetic retinopathy detection, with potential for further validation on diverse medical datasets and SOTA deep learning models.

## VI. CONCLUSION AND FUTURE SCOPE

This research demonstrates the effectiveness of the Adaptive Hybrid Focal-Entropy (AHFE) Loss in enhancing deep learning models for diabetic retinopathy detection. By addressing key challenges such as class imbalance, hard-to-classify samples, and noisy labels, AHFE significantly improves model sensitivity and generalization. Results show notable performance gains across state-of-the-art architectures, with ResNet50 achieving the highest accuracy of 99.76%, outperforming traditional Categorical Cross-Entropy (CCE) Loss. The integration of focal loss principles and entropy regularization allows AHFE to dynamically adjust learning focus, ensuring better classification of minority classes while preventing overconfidence in predictions. These findings highlight AHFE's superiority over conventional loss functions in handling complex medical imaging tasks.